\documentclass[runningheads]{svmult}

\usepackage{makeidx}   
\usepackage{graphicx}  
\usepackage{subeqnar}  
\usepackage{multicol}  
\usepackage{physprbb}  
\makeindex             

\begin{document}
%
\title*{Submillimeter Galaxies as Tracers of Mass Assembly at Large $M$}
\toctitle{Submillimeter Galaxies as Tracers of Mass Assembly at Large $M$}
%
%
\titlerunning{SMGS as Tracers of Mass Assembly at Large $M$}
%
\author{R. Genzel\inst{1,2}
\and A.~J. Baker\inst{1}
\and R.~J. Ivison\inst{3}
\and F. Bertoldi\inst{4}
\and A.~W. Blain\inst{5}
\and S.~C. Chapman\inst{5}
\and P. Cox\inst{6}
\and R.~I. Davies\inst{1}
\and F. Eisenhauer\inst{1}
\and D.~T. Frayer\inst{7}
\and T. Greve\inst{8}
\and M.~D. Lehnert\inst{1}
\and D. Lutz\inst{1}
\and N. Nesvadba\inst{1}
\and R. Neri\inst{9}
\and A. Omont\inst{10}
\and S. Seitz\inst{11}
\and I. Smail\inst{12}
\and L.~J. Tacconi\inst{1}
\and M. Tecza\inst{1}
\and N.~A. Thatte\inst{13}
\and R. Bender\inst{1,11}}
\authorrunning{R. Genzel et al.}
%
%
\institute{Max-Planck-Institut f{\" u}r extraterrestrische Physik (MPE),
Garching, Germany
\and Department of Physics, University of California, Berkeley, CA, USA
\and Astronomy Technology Centre, Royal Observatory, Edinburgh, UK
\and Max-Planck-Institut f{\" u}r Radioastronomie (MPIfR), Bonn, Germany
\and California Institute of Technology, Pasadena, USA
\and Institut d'Astrophysique Spatiale, Universit{\' e} de Paris Sud, Orsay, 
France
\and Spitzer Science Center, California Institute of Technology, Pasadena, USA
\and Institute for Astronomy, University of Edinburgh, Edinburgh, UK
\and Institut de Radio Astronomie Millim{\' e}trique (IRAM), Grenoble, France
\and CNRS \& Universit{\' e} de Paris, Paris, France
\and Munich University Observatory (USM), M{\" u}nchen, Germany
\and Institute for Computational Cosmology, University of Durham, Durham, UK
\and University of Oxford Astrophysics, Oxford, UK}

\maketitle              

\section{Introduction}\label{s-intro}
 
Deep imaging in the rest-frame UV has constrained both the evolution of the 
cosmic star formation rate density \cite{mada96} and its time integral, the 
growth of the cosmic stellar mass density \cite{dick03}.  Short-wavelength 
studies give an incomplete picture, however, since an important population of
high-redshift galaxies is heavily dust-obscured.  The strength of the 
extragalactic mid- and far-IR/submillimeter background indicates that about 
half of the cosmic energy density comes from dusty luminous and ultra-luminous 
infrared galaxies (LIRGs/ULIRGs: $L_{\rm IR} \sim 10^{11.5}$ to 
$10^{13.5}\,L_\odot$) at $z \geq 1$ \cite{puge96,pei99}.  Because the
brightest of these submillimeter galaxies (SMGs; see \cite{blai02} and
references therein) tend to lack strong X-ray emission \cite{alma03}, their 
large IR luminosities probably correspond to high star formation rates 
\cite{barg00}.  As the strikingly different appearances of the Hubble Deep 
Field at $0.83\,{\rm \mu m}$ \cite{will96} and $850\,{\rm \mu m}$ 
\cite{hugh98} exemplify, SMGs are rarer and forming stars much more intensely 
than typical optically selected systems \cite{will96,hugh98}.  Here we discuss
new observations that shed light on the importance of SMGs in the history of 
galaxy mass assembly (all numbers assuming a flat $\Omega_\Lambda = 0.7$ 
cosmology with $H_0 = 70\,{\rm km\,s^{-1}\,Mpc^{-1}}$).
                                       
\section{Observations: PdBI millimeter interferometry}\label{s-obs-mm}
 
\begin{figure}[ht]
\begin{center}
\includegraphics[width=0.8\textwidth]{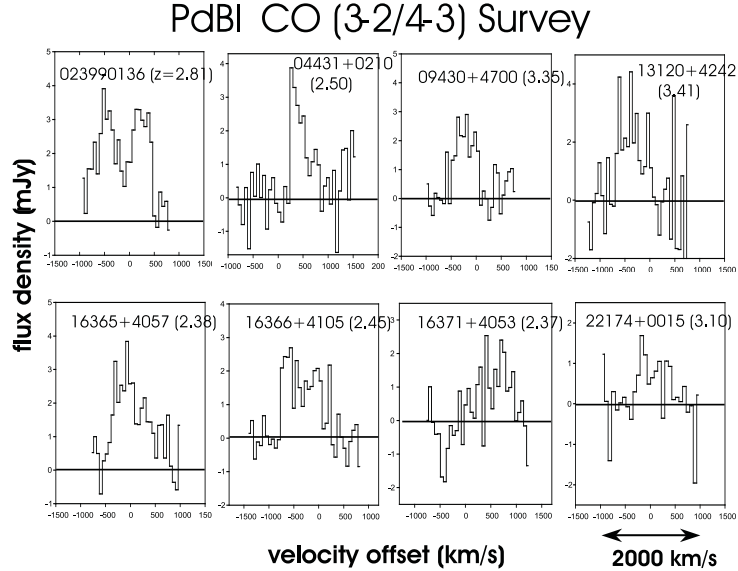}
\end{center}
\caption[]{Integrated CO spectra of the seven detected PdBI survey sources 
\cite{neri03,grev04} plus SMMJ\,02399-0136 \cite{genz03}.  Source names and 
redshifts are listed in the insets; flux density and velocity scales are the 
same for all sources.}
\label{eps1}
\end{figure}

\begin{figure}[ht]
\begin{center}
\includegraphics[width=0.8\textwidth]{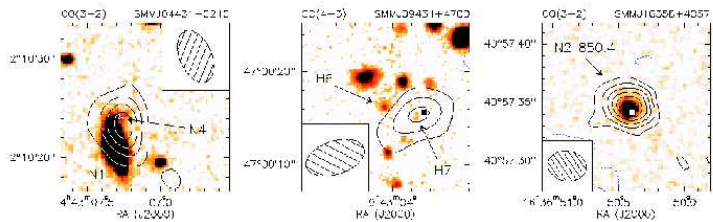}
\end{center}
\caption[]{Integrated CO maps of three SMGs studied with the PdBI 
\cite{neri03}, superposed on greyscale images of rest-frame UV (center panel)
and optical (left and right panels) emission.  FWHM synthesized beams are 
shown as hatched ellipses.  In the left panel, the asterisk marks the position 
of the ERO N4 (uncertainty $\pm 0.5^{\prime\prime}$ -- see text), while the 
edge-on spiral galaxy N1 ($2^{\prime\prime}$ SE of the CO source) lies in the
foreground cluster at $z=0.18$.  In the center and right panels, filled 
squares mark millimeter continuum positions.  Arrows show the positions of 
radio sources H6 and H7; the stronger of these (H6) is a narrow-line Seyfert 
galaxy with ${\rm Ly\,\alpha}$ indicating $z=3.349$. CO(4--3) emission from 
N2\,850.4 remains largely unresolved.}
\label{eps2}
\end{figure}

Since SMGs in most cases have relatively poorly known positions and frequently 
have only weak counterparts in the rest-frame UV and optical, redshifts and 
spectroscopic parameters were initially confirmed with CO interferometry 
for only two of the few hundred detected sources \cite{blai02}.  Recently, a 
subset of the authors have initiated a new program to obtain spectroscopic 
redshifts with the Keck telescope for a number of sources detected with SCUBA 
at $850\,{\rm \mu m}$, in most cases aided by more precise positions derived 
from deep 1.4\,GHz VLA maps \cite{chap03}.  Here we report the first results 
of the millimeter follow-up of these SMGs, undertaken with the IRAM Plateau 
de Bure Interferometer (PdBI).  In the B, C, and D configurations of the PdBI, 
we searched for the CO(3--2) and CO(4--3) rotational transitions (redshifted 
to the 3\,mm atmospheric window) in a sample of so far twelve SMGs with $S_{850} \geq 
5\,{\rm mJy}$ and reliable UV/optical spectroscopic redshifts.  Integration 
times per source ranged from 13 to 30 hours, for a total integration time of 
about 1Ms in the 2002/2003 observing season.  We have detected significant CO 
emission in 7 out of the 12 sources, thus roughly quadrupling the number of 
CO-confirmed SMG redshifts \cite{neri03,grev04}.  Figure 1 shows the 
integrated CO(3--2/4--3) spectra for the seven new sources and for the 
$z = 2.81$ system SMM\,J02399-0136 \cite{genz03}. Figure 2 shows the 
integrated CO maps superposed on rest-frame UV ($I$-band) and optical
($K$-band) images of three of the sources \cite{neri03}.

Our PdBI observations confirm the median $z \sim 2.4$ determined from 
rest-frame UV/optical spectroscopy \cite{chap03}, although in several 
cases, the UV redshift is inferred from a source that is physically distinct 
from the SMG (e.g., SMM\,J09431+4700 in Fig. 2).  SMGs likely sample very 
dense environments that may harbor additional, UV-bright star-forming galaxies.

SMGs are rich in molecular gas.  Correcting for foreground lensing wherever 
appropriate (and known), and using the $M_{\rm gas}/L_{\rm CO}$ conversion 
factor appropriate for $z \sim 0.1$ ULIRGs, the median gas mass of the eight 
SMGs in Fig. 1 (plus the $z=2.56$ galaxy SMM\,J14011+0252 
\cite{fray99,down03}) is 
$2.2 \times 10^{10}\,M_\odot$.  This value is similar to the median
molecular gas mass found in high-$z$ QSOs and about three times greater
than in local ULIRGs.  The inferred median far-IR luminosity of the sample
(again corrected for lensing) is about is $1.3 \times 10^{13}\,L_\odot$,
corresponding to a star formation rate of about $1100\,M_\odot\,{\rm yr^{-1}}$.
 
Our most important finding is that the brightest SMGs ($S_{850} \geq 5\,{\rm 
mJy}$), contributing about 25\% of the submillimeter background, have large 
dynamical masses. The median velocity dispersion of our sample is $255 (\pm 
38)\,{\rm km\,s^{-1}}$.  The ratio of the squares of the median velocity 
dispersions then implies that our SMGs have dynamical masses at least $13 (\pm 
4)$ times greater than the $z \sim 3$ Lyman break galaxy (LBG) population 
\cite{pett01}, and at least $5 (\pm 1.5)$ times greater than optically 
selected populations at $z \sim 2$ (\cite{erb03} -- see also this volume). 
There are indications that SMGs are larger than LBGs in which case these
estimates are lower limits \cite{genz03,neri03,chap04}. 
Assuming a rotating disk geometry at an inclination of $40^\circ$, the median 
velocity dispersion corresponds to a dynamical mass of $1.1 \times 
10^{11}\,M_\odot$ within $R=4\,{\rm kpc}$.  Based on its spatially resolved 
rotation curve, SMM\,J02399-0136 has a total dynamical mass of $3 \times 
10^{11}\,M_\odot$ within $R = 8\,{\rm kpc}$, half of which is probably 
stellar \cite{genz03}.  Near-IR photometry indicates that the median stellar 
mass for the $z \sim 3$ LBG population is $1.1 \times 
10^{10}\,M_\odot$\cite{shap01}, consistent with the dynamical mass ratio 
estimated above.  Dynamical and stellar masses of the SMGs in our sample are 
thus comparable to those of local $m^*$ galaxies.
 
\section{Observations: SPIFFI NIR integral field spectroscopy}
\label{s-obs-ir} 

\begin{figure}[ht]
\begin{center}
\includegraphics[width=0.7\textwidth]{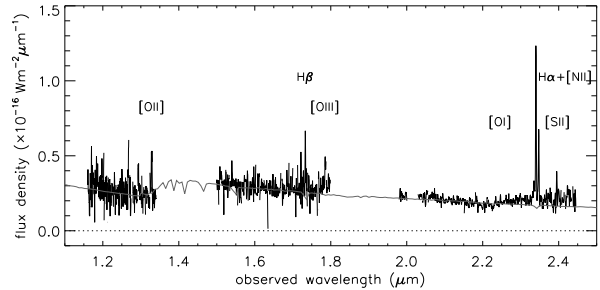}
\end{center}
\caption[]{$J$, $H$, and $K$ spectra of J14011 extracted from a 
$2^{\prime\prime} \times 1.5^{\prime\prime}$ aperture \cite{tecz04}.  
Superposed is the model STARS spectrum for a 200\,Myr old continuous star 
formation episode with solar metallicity, a $1-100\,M_\odot$ Salpeter 
IMF, and extinction $A_V = 0.7$.}
\label{eps3}
\end{figure}

We have also conducted a detailed rest-frame optical case study of the bright 
source SMM\,J14011+0252 (hereafter J14011), a $z = 2.565$ SMG lying behind the 
$z = 0.25$ cluster A1835 \cite{ivis01}.  For this study we used SPIFFI, the 
SPectrometer for Infrared Faint Field Imaging \cite{eise03}, at the ESO VLT.  
In three runs between February and April 2003, we observed J14011 in the $J$, 
$H$, and $K$ bands, with a FWHM spatial resolution of about 
$0.6^{\prime\prime}$ and spectral resolutions ranging from 125 to 
$200\,{\rm km\,s^{-1}}$.  SPIFFI obtains simultaneous spectra for each of 
$32 \times 32$ contiguous $0.25^{\prime\prime}$ pixels.  The total on-source 
integration times were 60 minutes in $J$, 95 minutes in $H$, and 340 minutes 
in $K$ \cite{tecz04}. 
 
Within the bright eastern J1 component of J14011, the SPIFFI data reveal 
substantial spatial variation in ${\rm H\alpha}$ equivalent width: nebular 
line emission is extended over about $1^{\prime\prime}$ (lensing corrected) 
and is strongest away from the central continuum peak.  The integrated
spectrum of J1 (Fig. 3) appears to exhibit a continuum break between the
observed $J$ and $H$ bands.  The feature's most likely origin is the Balmer 
break of a $\geq 100\,{\rm Myr}$ old stellar population at $z \simeq 2.565$.  
The J1 line fluxes allow us to measure the $R_{23}$ estimator for its oxygen 
abundance.  After correcting for the $E(B-V) = 0.18$ implied by the Balmer
decrement, we derive an oxygen abundance of $8.96 \pm 0.10$, or 1.9 times
solar. The observed high [NII]/${\rm H\alpha}$ ratio implies that J14011 
cannot fall on the lower-metallicity branch of the [O/H] vs. $R_{23}$ 
relation. In contrast to the high abundances claimed for some high-$z$ QSOs,
this metallicity refers to the entire galaxy on $\sim 10\,{\rm kpc}$ scales.
 
For a closed-box model with solar yield, this metallicity corresponds to
a gas/baryonic mass fraction of ${\rm exp}\,(-Z/Z_\odot) \simeq 0.16$.
From its intrinsic (corrected for a lensing magnification of 5)
gas mass of $\simeq 1.3 \times 10^{10}\,M_\odot$, we therefore infer a total 
baryonic mass $\simeq 8 \times 10^{10}\,M_\odot$ and stellar mass
$\simeq 7 \times 10^{10}\,M_\odot$. This last value is empirically 
supported by the location of J14011 in the local mass-metallicity
relation. Adopting a recent version of that relation \cite{trem04}, we 
would expect a stellar mass of $\sim 6 \times 10^{10}\,M_\odot$.  From the 
closed-box model with continuous star formation at the current rate 
($380\,M_\odot\,{\rm yr^{-1}}$) we infer the overall duration of the star 
forming activity in J14011 to be $t_{\rm SF} \simeq 220\,{\rm Myr}$.  This 
age is entirely consistent with the strength of the Balmer break seen in Fig. 
3.  As a further consistency check, we can estimate a dynamical mass for 
J14011 from the $\sim 2.2^{\prime\prime}$ and $\sim 180\,{\rm km\,s^{-1}}$ 
separations between J1 and [second component] J2, assuming these orbit around 
a common center.  For a predominantly north-south lensing shear, the result 
is $\sim 3.4 \times 10^{10}\,{\rm sin}^{-2}i\,M_\odot$, consistent with the 
moderately low inclination also implied by the ${\rm H\alpha}$ morphology.

\section {SMGs and mass assembly} \label{ss-massass}
 
\begin{figure}[ht]
\begin{center}
\includegraphics[width=0.6\textwidth]{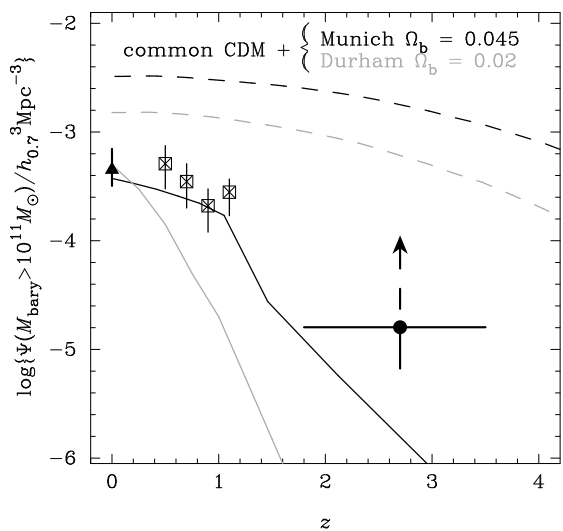}
\end{center}
\caption[]{Comoving number densities of galaxies with baryonic masses $\geq
10^{11}\,M_\odot$ as a function of redshift \cite{tecz04}.  Triangle and open 
squares show densities of massive stellar systems at $z = 0$ \cite{cole01}
and $z \sim 1$ \cite{dror03}; circle shows density for massive SMGs at 
$z \sim 2.7$, with a factor of 7 correction for burst lifetime.  Solid
curves show the predictions of semi-analytic modelling by the ``Munich'' and 
``Durham'' groups \cite{kauf99,baug03}; dashed curves show the corresponding 
number densities of halos with {\it available} baryonic masses 
$\geq 10^{11}\,M_\odot$.  The two models use the same halo simulations but 
assume different $\Omega_{\rm b}$.}
\label{eps4}
\end{figure}

Our IRAM and VLT results on metallicity and gas/dynamical mass build a 
compelling case that the most luminous SMGs ($S_{850} \sim 5$ to 10\,mJy)
have dynamical and -- within the factor 2 uncertainties of the present 
estimates -- also stellar masses of $\sim 10^{11}\,M_\odot$. This stellar 
component has plausibly formed in the current starburst of duration 
$\sim 100\,{\rm Myr}$. How does the inferred comoving volume density of SMGs 
compare to theoretical models of galaxy formation?
 
From the observed surface density of SMGs with $S_{850} > 5\,{\rm mJy}$ 
\cite{smai02}, we infer that radio-detected SMGs with baryonic masses $\geq 
10^{11}\,M_\odot$ in the range $1.8 \leq z \leq 3.5$ have a comoving number 
density of $1.6_{-0.6}^{+1.0} \times 10^{-5}\,{\rm Mpc}^{-3}$ \cite{tecz04}.
This density needs to be corrected upward to account for sources that
are in the same mass bin but no longer luminous enough to have been detected at
$850\,{\rm \mu m}$.  Relative to the 1.8\,Gyr elapsed over the range $1.8 
\leq z \leq 3.5$, the $\sim 220\,{\rm Myr}$ age and $\sim 30\,{\rm Myr}$ gas
exhaustion timescale for the starburst in J14011 imply a lifetime correction
factor of about 7.  Figure 4 plots this prediction,
together with the comoving number densities of galaxies with stellar masses
above the $10^{11}\,M_\odot$ threshold at $z \sim 0$ \cite{cole01} and $z
\sim 1$ \cite{dror03}.  Also plotted are the predictions of two semi-analytic
models \cite{kauf99,baug03}.  Since all theoretical and observational values
assume the same cosmology and a consistent (either Miller-Scalo or
$1-100\,M_\odot$ Salpeter) IMF, we are secure in concluding that the models
underpredict the number densities of massive galaxies at $z \sim 2.5$ 
\cite{tecz04}.  Couching Fig. 4 in terms of {\it mass} emphasizes that
the observed surface densities of SMGs cannot be explained using a very flat
IMF alone.  We also note that the models contain enough dark halos 
of total mass $\geq 10^{11}\,(\Omega_{\rm M}/\Omega_{\rm b})\,M_\odot$ to 
account for the observed comoving number densities of massive SMGs, provided 
all of their baryons can be rapidly and efficiently assembled into galaxies.  
SMGs thus impose a powerful ``baryonic mass assembly'' test at the upper end 
of the galaxy mass function \cite{genz03}.

Our correction of the observed number density of SMGs to account for their 
less dust-luminous descendants raises the question of what types of galaxies 
SMGs can plausibly evolve into.  Although LBGs are clearly excluded due to 
their smaller masses, an intriguing alternative is the newly identified
population whose red $J - K$ colors can stem from a strong Balmer break at
$z \sim 2.5$ \cite{fran03}.  These galaxies appear to be strongly clustered
\cite{dadd03} and -- assuming a uniform distribution in the redshift range
$2 \leq z \leq 3.5$ \cite{vand03} -- have a comoving number density $\sim 1.8
\times 10^{-4}\,{\rm Mpc^{-3}}$.  Relatively few sources with red $J-K$
colors are also luminous (sub)millimeter emitters, indicating that the two 
populations have little direct overlap.  With three examples in the HDF-S 
having (for a Miller-Scalo IMF) stellar masses $(0.6 - 1.4) \times 
10^{11}\,M_\odot$ \cite{sara04}, it would seem plausible that the more 
luminous objects with red $J-K$ colors at this epoch could have passed 
through a SMG phase at higher redshift.

%

\end{document}